\documentclass{webofc}
\usepackage[varg]{txfonts}
\usepackage{dsfont}

\DeclareMathAlphabet{\mathpzc}{OT1}{pzc}{m}{it}
\newcommand {\bb}[1]{\mbox{\boldmath $#1$}}
\newcommand{\ii}{\mathrm{i}}

\begin{document}
\title{Stationary and non-stationary solutions of the evolution \\ equation for neutrino in matter}

\author{\firstname{A. V.} \lastname{Chukhnova}\inst{1}\fnsep\thanks{\email{av.chukhnova@physics.msu.ru}} \and
        \firstname{A. E.} \lastname{Lobanov}\inst{1}\fnsep\thanks{\email{lobanov@phys.msu.ru}}
}

\institute{Department of Theoretical Physics, Faculty of Physics,
  Moscow State University, 119991 Moscow, Russia
          }

\abstract{
  We study solutions of the equation which describes the evolution of a neutrino propagating in dense homogeneous medium in the framework of the quantum field theory. In the two-flavor model the explicit form of Green function is obtained, and as a consequence the dispersion law for a neutrino in matter is derived. It is shown that there exist both the solutions describing
the stationary states and the solutions describing the spin-flavor coherent states of the neutrino. The stationary states may be different from the mass eigenstates, and the wave function of a state with a definite flavor should be constructed as a linear combination of the wave functions of the stationary states with coefficients, which depend on the mixing angle in matter. In the ultra-relativistic limit the wave functions of the spin-flavor coherent states coincide with the solutions of the quasi-classical evolution equation. Quasi-classical approximation of the wave functions of spin-flavor coherent states is used to calculate the probabilities of transitions between neutrino states with definite flavor and helicity.
}
\maketitle
\section{Neutrino wave equation} \label{sec-1}
The phenomenological theory of oscillations based on the ideas of B. Pontecorvo
\cite{Pontecorvo} and Z. Maki et al. \cite{MNS} describes the general properties of neutrino oscillations. As this theory was initially developed to describe neutrinos of rather high energies, it is not appropriate for study of low-energy neutrinos including the relic neutrinos, which play an important role in many cosmological models. So, for a complete description we need a theory, which is applicable for the low energy neutrinos and which gives the same results as the phenomenological theory for the high-energy neutrinos.

Such a description can be made on the base of a modification of the Standard Model, which was proposed in papers \cite{tmf2017,lobanov_osc}. In this modification, the fermions are combined in multiplets. One-particle wave functions are elements of the representation space of the direct product of the Poincar\'{e} group and the internal symmetry group $SU(3)$. Within this approach, the phenomenon of neutrino oscillations arises as a direct consequence of the general principles of quantum field theory. The probability of detecting a neutrino of a particular flavor may be calculated in the interaction picture, and the dependence of this probability on the distance between the source and the detector results in formulas for neutrino oscillations. In the ultra-relativistic limit these formulas are in a good agreement with those obtained in the phenomenological approach.

As it is well-known \cite{Wolf}, neutrinos propagating in matter interact with the medium via forward elastic scattering on background fermions. This interaction modifies the pattern of the flavor oscillations. In particular, it results in Mikheyev-Smirnov-Wolfenstein effect \cite{MS}, which can explain the deficit of solar neutrinos \cite{Bethe_1986}. If the medium is moving or if it is polarized, then the spin oscillations also occure \cite{Lobanov_5}.

An effective equation, which describes both the neutrino oscillations and the spin rotation,  arises due to reduction of the mass operator of neutrino in medium \cite{izvu_2016}. Therefore, there is an upper limit on neutrino energies, for which the equation is applicable. However, the equation may be used for neutrinos of arbitrary low energies. We consider the medium consisting of neutrons, protons and electrons. In this case the evolution equation takes the form
\begin{equation}\label{i1}
\bigg(i\gamma^{\mu}{\partial}_{\mu} - {\mathds{M}}-
\frac{1}{2}
\gamma^{\mu}f^{(e)}_\mu (1 + \gamma^5){\mathds P}^{(e)} -\frac{1}{2}
\gamma^{\mu}f^{({\mathrm N})}_{\mu}(1 + \gamma^5)\,{\mathds I} \!\bigg)\,{\Psi}(x) = 0.
\end{equation}
\noindent  Here ${\mathds{I}}$ is a $3\times 3$ identity matrix, ${\mathds{M}}$ is a Hermitian mass matrix of the neutrino multiplet, which can be written as follows
\begin{equation}\label{s1-2}
{\mathds{M}}=
\sum\limits_{l=1}^{3}{m_{l}}{\mathds{P}}^{(l)},
\end{equation}
\noindent where $m_{l}$ are the eigenvalues of the mass matrix, which have the meaning of the masses of the multiplet components, and the matrices ${\mathds{P}}^{(l)}$ are orthogonal projectors on the subspaces of wave functions, which describe the states with these masses. The matrix ${\mathds{P}}^{(e)}$ is a projector on the state of neutrino with electron flavor. In Eq. \eqref{s1-2} the product of the Dirac matrices and the matrices $\mathds{M}$, ${\mathds{P}}^{(e)}$ is defined as tensor product.

In the lowest order of the perturbation theory the effective potentials describing the interaction of the neutrino with the medium via charged currents $f^{\mu (e)}$ and via neutral currents $f^{\mu (\mathrm N)}$ are as follows
\begin{equation}\label{l12}
f^{\mu(e)} =\sqrt{2}{G}_{{\mathrm F}}\left({j^{\mu (e)}}
-\lambda^{\mu(e)}\right),
\end{equation}
\begin{equation}\label{l14}
f^{\mu  (\mathrm N)} =\sqrt{2}{G}_{{\mathrm F}}\sum\limits_{i=e,p,n}
\left({j^{\mu (i)}}
\left(T^{(i)}-2Q^{(i)}\sin^{2}
\theta_{\mathrm{W}}\right)-
{\lambda^{\mu(i)}}T^{(i)}\right).
\end{equation}
\noindent Here $j^{\mu (i)}$ are the $4$-vectors of currents and $\lambda^{\mu (i)}$ are the $4$-vectors of polarization of the background fermions. In these formulas,  $T^{(i)}$ are the projections of the weak isospin and $Q^{(i)}$ are the electric charges of the fermions of the medium, $G_{\mathrm F}$ is the Fermi constant, $\theta_{\mathrm W}$ is the Weinberg angle.

In the three-flavor model the wave function $\Psi(x)$ is a twelve-component object. It is convenient to introduce block structure, namely to define this object using three Dirac bispinors
\begin{equation}
\Psi(x) = \left( \begin{matrix} \psi_1(x) \\ \psi_2(x) \\ \psi_3(x) \end{matrix} \right).
\end{equation}
In this case the $\gamma$-matrices act on the components of the Dirac bispinors and the matrices $\mathds{M}$, $\mathds{P}^{(e)}$ permute the bispinors $\psi_i (x)$. In the two-flavor model the wave function $\Psi(x)$ has eight components.

As in the case of the $\gamma $-matrices in the Dirac equation, the matrices $\mathds{M}$ and $\mathds{P}^{(e)}$ can be written in different representations, which are connected by unitary transformations.
We will introduce the mass representation of the matrices $\mathds{M}$,  $\mathds{P}^{(e)}$ as a representation, in which the mass matrix is diagonal.
We also introduce the flavor representation of these matrices as a representation, in which the flavor projectors are diagonal matrices.
These representations are connected by the Pontecorvo-Maki-Nakagawa-Sakata mixing matrix $U_{PMNS}$.

Any solution of Eq. \eqref{i1} is a wave function describing a state of the neutrino. The mass states are the states described by the wave functions $\Psi_i(x)$ ($i = 1,2,3 $), which can be written in the mass representation at every point of the event space in the following form
\begin{equation}\label{WF_m}
\Psi_1(x) = \left( \begin{matrix} \psi_1(x) \\ 0 \\ 0 \end{matrix} \right), \qquad
\Psi_2 (x) = \left( \begin{matrix} 0\\ \psi_2(x) \\ 0 \end{matrix} \right), \qquad
\Psi_3 (x) = \left( \begin{matrix} 0 \\ 0\\ \psi_3(x) \end{matrix} \right),
\end{equation}
where $\psi_i(x)$ ($ i=1,2,3 $) are Dirac bispinors.

The neutrino is in a state with a definite flavor at a certain space-time point if its wave function $\tilde{\Psi}_i(x)$ ($i = 1,2,3$) in the flavor representation takes the form
\begin{equation}\label{WF_f}
\tilde{\Psi}_1(x) = \left( \begin{matrix} \tilde{\psi}_1 (x)\\ 0 \\ 0 \end{matrix} \right), \qquad
\tilde{\Psi}_2(x) = \left( \begin{matrix} 0\\ \tilde{\psi}_2(x)\\ 0 \end{matrix} \right), \qquad
\tilde{\Psi}_3(x) = \left( \begin{matrix} 0 \\ 0\\ \tilde{\psi}_3(x) \end{matrix} \right),
\end{equation}
where $\tilde{\psi}_i(x)$ ($i=1,2,3$) are Dirac bispinors.

\section{Green function} \label{sec-2}
We will investigate the neutrino propagation in the homogeneous matter. If the medium is non-polarized and all its components are moving with the same velocity, then the effective potentials of interaction with the medium via charged currents and via neutral currents are proportional
\begin{equation}
f^{\mu (e)}=f^\mu, \qquad f^{\mu (\mathrm N)} = a f^\mu.
\end{equation}
\noindent The proportionality factor $a$ depends on the number densities $n^{(i)}$ of the background fermions as follows
 \begin{equation} \label{a}
 a=\sum\limits_{i=e,p,n}
\frac{n^{(i)}}{n^{(e)}}(T^{(i)}-2Q^{(i)}\sin^{2}
\theta_{\mathrm{W}} ).
\end{equation}
\noindent In this case Eq. \eqref{i1} takes the form
\begin{equation}\label{i3}
\left( \ii \partial^\mu \gamma_\mu - \frac{1}{2} \gamma^\mu f_\mu (1+\gamma^5) (a\mathds{I}+
\mathds{P}^{(e)}) - \mathds{M}\right)\Psi(x) = 0.
\end{equation}

In the momentum representation Green function of Eq. \eqref{i3} is the inverse operator of the equation
\begin{equation}\label{g0}
G(p) = \left({  \gamma^\mu p_\mu\mathds{I} - \frac{1}{2}  \gamma^\mu f_\mu(1+\gamma^5)(a\mathds{I}+\mathds{P}^{(e)}) - \mathds{M}}\right)^{-1}.
\end{equation}
We introduce the following notation
\begin{equation}\label{g1}
H_{+}(p) = \left({ \gamma^\mu p_\mu \mathds{I} - \frac{1}{2}\gamma^\mu f_\mu (1+\gamma^5)(a\mathds{I}+\mathds{P}^{(e)}) + \mathds{M}}\right).
\end{equation}

\noindent It can be shown that in the two-flavor model
\begin{equation}\label{g10}
G(p) =
\frac{1}{2}\sum\limits^{}_{\zeta=\pm 1}\frac{H_{+}(p,\zeta)F_{-}(p,\zeta)}{D(p,\zeta)}(1+\zeta {\cal S}).
\end{equation}
Here
\begin{multline}\label{g8}
F_{-}(p,\zeta) = p^2 - ((pf)-R\zeta)\left(a+\frac{1}{2}-\frac{1}{2}\,\sigma_{3}\right)- \frac{m_1^2+m_2^2}{2}-\\  -\frac{m_2^2-m_1^2}{2}(\sigma_3 \cos{2\theta} - \sigma_1\sin{2\theta})- \frac{\ii}{4} \gamma^\mu f_\mu(1+\gamma^5)(m_2-m_1)\sigma_2\sin{2\theta},
\end{multline}
\begin{multline} \label{disp}
D(p,\zeta) = \left( p^2 - ((pf)-R\zeta)\left(a+\frac{1}{2}\right)- \frac{m_1^2+m_2^2}{2}\right)^{2}-\\ - \left(\frac{((pf)-R\zeta)}{2}-\frac{m_2^2-m_1^2}{2} \cos{2\theta}\right)^{2}- \left(\frac{m_2^2-m_1^2}{2}\sin{2\theta})\right)^{2},
\end{multline}
where $\sigma_{i}$ are the Pauli matrices and $\zeta=\pm 1$ are the eigenvalues of the operator ${\cal S}$ (${\cal S}^{2}=1$), which describes the projection of neutrino spin on its canonical momentum in the rest frame of the medium.
In the reference frame, where the medium is moving, it takes the form
\begin{equation}\label{g4}
{\cal S}= \frac{1}{2{R}}\gamma^5({\gamma^\mu f_\mu   \gamma^\nu p_\nu-\  \gamma^\nu p_\nu \gamma^\mu f_\mu}),
\end{equation}
where $R = \sqrt{(fp)^{2}-f^{2}p^{2}}$.

The poles of the denominator of the Green function, i. e. the solutions of the equation
\begin{equation} \label{pol}
D(p,\zeta) = 0,
\end{equation}
\noindent correspond to the eigenvalues of the Hamiltonian. That is, Eq. \eqref{pol} defines the dispersion law for neutrino. This equation is an equation of degree four. It remains unchanged if we change the sign of the the $4$-momentum and the effective potential. Two of the solutions of Eq. \eqref{pol} correspond to neutrino energies and two correspond to antineutrino energies
\begin{equation} \label{disp2}
D(p,\zeta) =(p^{0}-\varepsilon_{+}({\bf p},\zeta))(p^{0}-\varepsilon_{-}({\bf p},\zeta))
(p^{0}+\bar{\varepsilon}_{+}({\bf p},\zeta))(p^{0}+\bar{\varepsilon}_{-}({\bf p},\zeta)),
\end{equation}
\noindent  where $\varepsilon_{\pm}$ are the neutrino energies, and $\bar{\varepsilon}_{\pm}$ are antineutrino energies.
Meanwhile, the neutrino and the antineutrino energies are connected by the transformation $\bar{\varepsilon}_{\pm}
({\bf p},\zeta, f) =\varepsilon_{\pm}(-{\bf p},\zeta,-f)$.

\section{Stationary solutions} \label{sec-4}

Using the equation for the Green function, we can get a stationary basis of the space of solutions of Eq. \eqref{i3}. The elements of the basis for the positive-frequency solutions are as follows
\begin{equation}\label{g13}
\varPsi_{{\bf p},\zeta}^{\pm}(x)=\int dp^{0}e^{-\ii(px)}\delta(p^{0}-
\varepsilon_{\pm}({\bf p},\zeta))\,\frac{(p^{0}-\varepsilon_{\pm}({\bf p},
\zeta))H_{+}(p,\zeta)F_{-}(p,\zeta)}{D(p,\zeta)}\,\varPsi_0(\zeta).
\end{equation}
\noindent  Here
\begin{equation}\label{g14}
\varPsi_0(\zeta)=\psi(\zeta)\otimes e_{\pm}, \quad \bar{\varPsi}_0(\zeta)
\varPsi_0(\zeta)= 1,
\end{equation}
\noindent  where
$\psi(\zeta)$ is a bispinor, which is an eigenvector of the spin projector operator ${\cal S}$ with the eigenvalue $\zeta$, and
$e_{\pm}$ are arbitrary orthogonal  unit vectors in a two-dimensional space over the field of complex numbers. By analogy with Eq. \eqref{g13}, we can write the elements of the basis for negative-frequency solutions.

Unfortunately, the expression \eqref{g13} is rather cumbersome.
To study the explicit form of the solutions we consider the neutrino moving in non-polarized matter at rest, when
$f^{\mu}=\{f_{0}, 0,0,0\}$. In this case the spin projection operator $\cal{S}$
does not depend on the effective potentials and it coincides with the helicity operator
\begin{equation}\label{op}
{\cal S} = \gamma^5 \frac{\gamma^{\mu}f_{\mu}\gamma^{\nu}p_{\nu}-\gamma^{\nu}p_{\nu}\gamma^{\mu}f_{\mu}}{2R}\rightarrow
\frac{(\bf \Sigma  p)}{|{\bf p}|}, \quad {\bf\Sigma} =
\gamma^5{\bb\gamma} \gamma^0.
\end{equation}
\noindent  That is, the helicity of the neutrino moving in dense matter is conserved. As a consequence, the form of the solutions simplifies.

Stationary solutions are the solutions of the following form
\begin{equation}\label{spinor}
\Psi(x) = e^{-i(px)}\Psi_0(p),
\end{equation}
where $\Psi_0(p)$ does not depend on the coordinates of the event space. Using the standard representation of the $\gamma$-matrices, we can write $\Psi_{0}(p)$
 for the neutrino with definite helicity as follows
\begin{equation} \label{s5}
\Psi_{0}(p) = \left(\begin{matrix} A_1^\zeta \chi_{\zeta}\\
A_2^\zeta\chi_{\zeta}\\ B_1^\zeta\chi_{\zeta} \\ B_2^\zeta\chi_{\zeta} \end{matrix} \right).
 \end{equation}
where $\chi_\zeta$ is a three-dimensional spinor, which describes a neutrino with the spin oriented along the direction of the canonical momentum
\begin{equation} \label{s05}
\chi_{\zeta}= \left(\begin{matrix} \chi^1_{\zeta}  \\ \chi^2_{\zeta} \end{matrix} \right),\qquad
 \frac{({\bb \sigma}{\bf p})}{|{\bf p}|}\chi_{\zeta} = \zeta \chi_{\zeta}.
\end{equation}

The equation on $A^\zeta_1$,  $A^\zeta_2$, $B^\zeta_1$, $B^\zeta_2$ has non-trivial solutions when
\begin{equation} \label{disp3}
(p^0- \zeta|{\bf p}|)^2 + 2(\zeta |{\bf p}| - \frac{f_0}{2}(a+1/2))(p^0- \zeta|{\bf p}|) - \frac{m_1^2+m_2^2}{2} = \frac{\xi}{2}\Delta ,
\end{equation}
where
\begin{equation}
\Delta = \sqrt{(f_0(p^0- \zeta|{\bf p}|)-(m_1^2-m_2^2) \cos{2\theta})^2 + (m_1^2-m_2^2)^2 \sin^2{2\theta}}\, ,
\end{equation}
\noindent and $\xi$ takes the values either $1$ or $-1$. In vacuum these values of $\xi$ correspond to the energies of different mass states. Coefficients $A^\zeta_1$,  $A^\zeta_2$, $B^\zeta_1$, $B^\zeta_2$ are defined up to a multiplicative constant $N$
\begin{eqnarray}\label{n1}
N p^0 A_1^\zeta &=& \frac{f_0 \sin{2\theta}}{4}(p^0-\zeta |{\bf p}| + m_1)(p^0 - \zeta |{\bf p}| - m_2), \\
N p^0 A_2^\zeta &=& - \frac{f_0 \sin{2\theta}}{4}(p^0-\zeta |{\bf p}| - m_1)(p^0 - \zeta |{\bf p}| - m_2), \\
N p^0 B_1^\zeta &=& \frac{f_0^2 \sin^2{2\theta}}{8}(p^0- \zeta |{\bf p}|) + \frac{1}{2}\left(\zeta |{\bf p}| - \frac{f_0}{2}\left(a+\frac{1}{2}-\frac{\cos{2\theta}}{2}\right)\right) \notag\\
&&\times \left( m_1^2-m_2^2-\xi \Delta  + f_0 (p^0-\zeta p) \cos{2\theta}\right),\\
N p^0 B_2^\zeta &=& \frac{f_0^2 \sin^2{2\theta}}{8}(p^0- \zeta |{\bf p}|) + \frac{1}{2}\left(p^0- m_2 - \frac{f_0}{2}\left(a+\frac{1}{2}-\frac{\cos{2\theta}}{2}\right)\right)  \notag\\  &&\times \left(m_1^2-m_2^2-\xi \Delta +  f_0(p^0-\zeta |{\bf p}|) \cos{2\theta}\right).\label{n2}
\end{eqnarray}
 The constant $N$ is chosen in order to satisfy the normalization condition
\begin{equation}\label{s007}
(A_1^\zeta)^2+(A_2^\zeta)^2+(B_1^\zeta)^2+(B_2^\zeta)^2=1.
\end{equation}

For ultra-relativistic neutrino $p^0 \gg m_i$ in the medium with reasonable density,  when $f_0 \ll p^0$, the formulas \eqref{n1}-\eqref{n2} become more simple. In this case the solutions for left-handed neutrinos ($\zeta = -1$) in the mass representation are as follows
 \begin{equation}\label{s8}
\Psi_+(x) = \frac{1}{\sqrt{2}}\left( \begin{matrix} \chi_{-}\sin{\phi}_{+}\\
-\chi_{-}\sin{\phi}_{+} \\ \chi_{-}\cos{\phi}_{+}
\\-\chi_{-}\cos{\phi}_{+} \end{matrix} \right)
e^{-\ii(\varepsilon_+ t - {\bf p x})},
\quad
\Psi_-(x) = \frac{1}{\sqrt{2}}\left( \begin{matrix} \chi_{-}\cos{\phi}_{-}
\\-\chi_{-}\cos{\phi}_{-} \\-\chi_{-}\sin{\phi}_{-}
\\ \chi_{-}\sin{\phi}_{-} \end{matrix} \right)
e^{-\ii(\varepsilon_- t - {\bf p x})},\end{equation}
Here $\varepsilon_+$, $\varepsilon_-$ are the energies corresponding to $\xi=+1$ and $\xi = -1$ in \eqref{disp3}. In these formulas we use the notations
\begin{equation}\label{phi}
\begin{array}{l}
\sin2{\phi}_{\pm} = \displaystyle \frac{2f_0p^0 \sin{2\theta}}{\sqrt{(2f_0p^0\sin{2\theta} )^2+
(2f_0 p^0\cos{2\theta}-(m_2^2-m_1^2))^2}},\\
 \cos2{\phi}_{\pm} =\displaystyle  \frac{(m_2^2-m_1^2) - 2f_0p^0 \cos{2\theta}}
{\sqrt{(2f_0p^0\sin{2\theta})^2+ (2f_0p^0 \cos{2\theta}-(m_2^2-m_1^2))^2}},
\end{array}
\end{equation}
\noindent where $p^0=\varepsilon_{\pm}$. To satisfy the orthogonality condition with the accuracy chosen we have to assume that
 $p^{0}=(\varepsilon_{+}+\varepsilon_{-})/2$ and ${\phi}_{+}={\phi}_{-}=\phi$.

In the limit $f_0 \rightarrow 0$ the solutions of the evolution equation are the wave functions of the mass states in vacuum. Then $\varepsilon_+$ corresponds to the second mass state and $\varepsilon_-$ corresponds to the first mass state. However, in matter that is not the case. So, the mass states of neutrino in medium do not exist.

In the flavor representation the stationary solutions with the canonical momentum ${\bf p}$ for $\xi=1$ and $\xi=-1$ take the form
\begin{equation}\label{s10}
\tilde{\Psi}_+(x) =\frac{1}{\sqrt{2}}\left( \begin{matrix} \chi_{-}\sin{\theta_{m}}
\\-\chi_{-}\sin{\theta_{m}}
\\ \chi_{-}\cos{\theta_{m}}  \\-\chi_{-}\cos{\theta_{m}}
 \end{matrix}\right) e^{-\ii(\varepsilon_+ t - {\bf p x})},
\quad
\tilde{\Psi}_-(x) = \frac{1}{\sqrt{2}}\left( \begin{matrix} \chi_{-}\cos{\theta_{m}}
\\-\chi_{-}\cos{\theta_{m}}
\\-\chi_{-}\sin{\theta_{m}}   \\ \chi_{-}\sin{\theta_{m}}
 \end{matrix}\right) e^{-\ii(\varepsilon_- t - {\bf{px}})},
\end{equation}
where $\theta_{m} = \theta + \phi$ is the effective mixing angle in matter. It is well-known \cite{Wolf}, that $\theta_{m}$ is defined as follows
\begin{equation}\label{dd}
\begin{array}{l}
\sin2\theta_{m} = \displaystyle \frac{(m_2^2-m_1^2) \sin{2\theta}}
{\sqrt{(2f_0p^0\sin{2\theta} )^2+
(2f_0 p^0\cos{2\theta}-(m_2^2-m_1^2))^2}},\\
 \cos2\theta_{m} =\displaystyle  \frac{(m_2^2-m_1^2) \cos{2\theta} - 2f_0 p^0}
{\sqrt{(2f_0p^0\sin{2\theta})^2+ (2f_0p^0 \cos{2\theta}-(m_2^2-m_1^2))^2}}.
\end{array}
\end{equation}
Thus, we can construct the wave functions of a neutrino with a definite flavor as a linear combination of the wave functions of the stationary states. The coefficients in this linear combination depend on the effective mixing angle in matter.

Right-handed neutrinos are described by the vacuum wave functions, where $\chi_{-}$ should be changed to $\chi_{+}$.

The dispersion law \eqref{disp3} generalizes the results of
 \cite{5,6,7,10} on the two-flavor case. However, it should be noted that the dispersion law \eqref{disp3} for neutrino propagating in non-polarized matter at rest was written yet in \cite{Kiers_1997}, though in that paper it was not consistently derived.

\section{Wave functions of spin-flavor coherent states } \label{sec-5}
However, there are solutions of another interesting type, namely the wave function of a neutrino with definite velocity. Indeed, we can find the solutions characterized not by a definite value of the canonical momentum, but by a definite value of the kinetic momentum $q^\mu$, which is connected with the $4$-velocity of the neutrino $u^\mu$ by the relation $q^\mu = m u^\mu$. Hence, $q^2=m^2$, where $m$ is determined  by the vacuum expectation value of the Higgs boson. We will call such solutions the wave functions of spin-flavor coherent states.

These solutions can be written with the help of a resolvent $\tilde{U}(x)$ \cite{SpinLight,arlomur} in the form
\begin{equation}\label{sf-5}
 \Psi(x)=\frac{1}{\sqrt{2 q_0}} \tilde{U}(x)\Psi_{0},
\end{equation}
\noindent where
\begin{equation}\label{sf2}
\Psi_0=\frac{1}{2}(1-\gamma^{5}\gamma_{\mu}{s}_{0}^{\mu}
)(\gamma_\mu q^\mu+m)\left(\psi^0\otimes e_{j}\right), \quad \bar{\Psi}_0\Psi_0= 2m.
\end{equation}
\noindent Here  $\psi^0$ is a constant bispinor,
 $e_{j}$ is an arbitrary unit vector in the three-dimensional vector space over the field of complex numbers, ${s}_{0}^{\mu}$ is the 4-vector of neutrino polarization, which satisfies the condition $(q{s}_{0})=0$.

The expression for resolvent is rather intricate. Therefore, we will restrict ourselves to the case of neutrinos of very high energies.
In the ultra-relativistic limit, we can use the quasi-classical description of the neutrino if we consider
 $x^\mu$ as a coordinate of the particle instead of the coordinate of the event space. So, $x^\mu = \tau q^\mu /m$, which means that the neutrino multiplet has a definite velocity $u^\mu=q^\mu/m$.
The proper time $\tau$ is connected with the distance $L$ between the source and the detector by the relation
$\tau = mL/|{\bf q}|$.

Using this approximation,
in the two-flavor model in the flavor representation we have
\begin{multline}\label{sf07}
 \tilde{U}(\tau) = \frac{1}{2}\sum\limits_{\zeta = \pm 1} \exp{\left\{ - \frac{\ii}{2}\tau
 \left( (m_2+m_1)
+ \big( (fu) - \zeta {\textstyle\sqrt{(fu)^{2}-f^{2}u^{2}}}\,  \big)\left(a+\frac{1}{2}\right)
\right)\right\}} \\ \times
\left( \cos(\tau\tilde{Z}_\zeta/{2}) - \ii \sin(\tau \tilde{Z}_\zeta /{2})
\left( \tilde{X}_\zeta \sigma_1 - \tilde{Y}_\zeta \sigma_3 \right) \right)(1-\zeta\gamma^{5}\gamma_{\mu}{s}^{\mu}),
 \end{multline}
\noindent where
\begin{equation}\label{sf7}
\begin{array}{l}
\tilde{Y}_\zeta = {\displaystyle \frac{1}{Z_\zeta}} \Big( (m_2-m_1)\cos{2\theta}
 -\big( (fu) -\zeta \sqrt{(fu)^{2}-f^{2}u^{2}}\, \big)/2 \Big),\\[8pt]
 \tilde{X}_\zeta = {\displaystyle \frac{1}{Z_\zeta}} \Big( (m_2-m_1)\sin{2\theta}
  \Big),\\ [8pt]
\tilde{Z}_\zeta=\sqrt{ \Big(\big( (fu)-\zeta \sqrt{(fu)^{2}-f^{2}u^{2}}\, \big)
/2 -(m_2-m_1)\cos{2\theta}\Big)^{2}+\Big((m_2-m_1)\sin{2\theta}
\Big)^2}.
\end{array}
\end{equation}
The solution  determined  by  resolvent \eqref{sf07}
coincides with the solution of the quasi-classical evolution equation, which was obtained in  \cite{VMU2017}.


\section{Quasi-classical approach for transition probabilities} \label{sec-6}

Using the quasi-classical limit of the wave functions of the spin-flavor coherent states, we can calculate the probabilities of the transitions between states $\alpha$ and $\beta$ with definite flavor and helicity. It is important to consider such states, since only left-handed neutrinos of the electron flavor are registered experimentally. In the two-flavor model, the projectors on the states with definite flavors take the following form
\begin{equation}\label{tt5}
{\mathds P}_{0}^{(\alpha)}=\frac{1}{2}\left(1+\xi_{0}\sigma_{3}\right),
\quad{\mathds P}_{0}^{(\beta)}=\frac{1}{2}\left(1+\xi'_{0}\sigma_{3}\right),
\quad\xi_{0},\xi'_{0}=\pm 1.
\end{equation}
Here $\xi_{0},\xi'_{0}=+ 1$ correspond to the electron flavor of the neutrino.
\noindent The states with a definite helicity are described by the $4$-vectors of the neutrino polarization
\begin{equation}\label{t6}
{s}_{0}^{(\alpha)\mu}=\zeta_{0}{s}^{\mu}_{sp},\quad {s}_{0}^{(\beta)\mu}=\zeta'_{0}{s}^{\mu}_{sp},\quad{s}^{\mu}_{sp}=
\{|{\bf u}|,u^{0}{\bf u}/|{\bf u}|\}.
\end{equation}
\noindent Here $\zeta_0,\zeta_0'= +1$ correspond to right-handed neutrino and $\zeta_0,\zeta_0'= -1$ correspond to left-handed neutrino.

The transition probability is as follows
\begin{multline}\label{l31}
W_{\alpha \rightarrow\beta}=\frac{1+\xi_0\xi'_0}{2}\frac{1+\zeta_0\zeta_0'}
{2}W_1 + \frac{1+\xi_0\xi'_0}{2}\frac{1-\zeta_0\zeta_0'}{2}W_2
 \\ + \frac{1-\xi_0\xi'_0}{2}\frac{1+\zeta_0\zeta_0'}{2}W_3 + \frac{1-\xi_0\xi'_0}{2}\frac{1-\zeta_0\zeta_0'}{2}W_4,
\end{multline}
\noindent where
\begin{equation}\label{ll32}
\begin{array}{l}
\displaystyle W_1=\frac{1}{2}\bigg(\frac{1}{2}
(1-\zeta_0(s s_{sp}))^2 (1- S_{+1}^2 X_{+1}^2) +\frac{1}{2}(1+\zeta_0(s s_{sp}))^2(1-
S_{-1}^2 X_{-1}^2) \\ [8pt] \displaystyle {\phantom{W_1=}}+(1- (s s_{sp})^2)
(C_{+1} C_{-1}+ S_{+1} S_{-1}
Y_{+1} Y_{-1})\cos({\omega\tau})  \\ \displaystyle {\phantom{W_1=}}
+ \xi_0(1-(s s_{sp})^2)(S_{+1} Y_{+1} C_{-1}- C_{+1} S_{-1}Y_{-1})\sin({\omega\tau }) \bigg), \\ \displaystyle
W_2=\frac{1}{2}\bigg( \frac{1}{2}(1-(s s_{sp})^2)(2- S_{+1}^2 X_{+1}^2- S_{-1}^2 X_{-1}^2)
   \\ [8pt] \displaystyle {\phantom{W_1=}}- (1-(s s_{sp})^2)(C_{+1} C_{-1} + S_{+1} S_{-1} Y_{+1}Y_{-1})\cos({\omega\tau }
  )   \\ \displaystyle {\phantom{W_1=}}-
\xi_0(1-(s s_{sp})^2)(S_{+1} Y_{+1} C_{-1}- C_{+1} S_{-1} Y_{-1})\sin({\omega\tau }) \bigg), \\
\displaystyle
W_3= \frac{1}{2}\bigg( \frac{1}{2}(1-\zeta_0(s s_{sp}))^2 S_{+1}^2 X_{+1}^2+ \frac{1}{2}(1+\zeta_0(s s_{sp}))^2 S_{-1}^2 X_{-1}^2 \\ [-4pt] \displaystyle {\phantom{W_1=}}+ (1-(s s_{sp})^2)S_{+1} S_{-1}X_{+1} X_{-1}\cos({\omega\tau}) \bigg), \\ \displaystyle
W_4= \frac{1}{2}\bigg( \frac{1}{2}(1-(s s_{sp})^2)(S_{+1}^2 X_{+1}^2
+ S_{-1}^2 X_{-1}^2)\\ [-4pt]\displaystyle {\phantom{W_1=}} - (1-(s s_{sp})^2)S_{+1} S_{-1}X_{+1} X_{-1}\cos({\omega\tau}) \bigg).
\end{array}
\end{equation}
\noindent  In these formulas,
\begin{equation}\label{ff}
\begin{array}{lll}
 C_{\pm 1}&=& \cos\big({\tau Z_{\pm 1}}/2\big),\quad
S_{\pm 1}=\sin\big({\tau Z_{\pm 1}}/2\big), \\[6pt]
 \omega &=& (1/2+a){\sqrt{(fu)^{2}-f^{2}}},\quad
\displaystyle  {s}^{\mu}= \frac{u^{\mu}(fu)
  - f^{\mu}}{\sqrt{(fu)^{2}-f^{2}}}, \\[8pt]
 X_{\pm 1}&=&\displaystyle\Big( \big(m_{2}-m_{1}\big)\sin 2\theta\Big)/Z_{\pm 1},\\
Y_{\pm 1}&=&\displaystyle\Big(\big( (fu) \mp {\sqrt{(fu)^{2}-f^{2}}}\big)/2 - \big(m_{2}-m_{1}\big)\cos2\theta\Big)/Z_{\pm 1}, \\[8pt]
Z_{\pm 1}&=&\left\{\Big(\big( (fu) \mp  {\sqrt{(fu)^{2}-f^{2}}}\big)/2 - \big(m_{2}-m_{1}\big)\cos2\theta\Big)^{2}+\Big(\big(m_{2}-m_{1}\big)
\sin 2\theta\Big)^{2}\right\}^{1/2}.
\end{array}
\end{equation}
\noindent These probabilities depend on six frequencies.

\begin{figure}
\vskip-20pt
\begin{minipage}[H]{0.47\linewidth}
\includegraphics[width=\linewidth, viewport=60 495 240 745,clip]{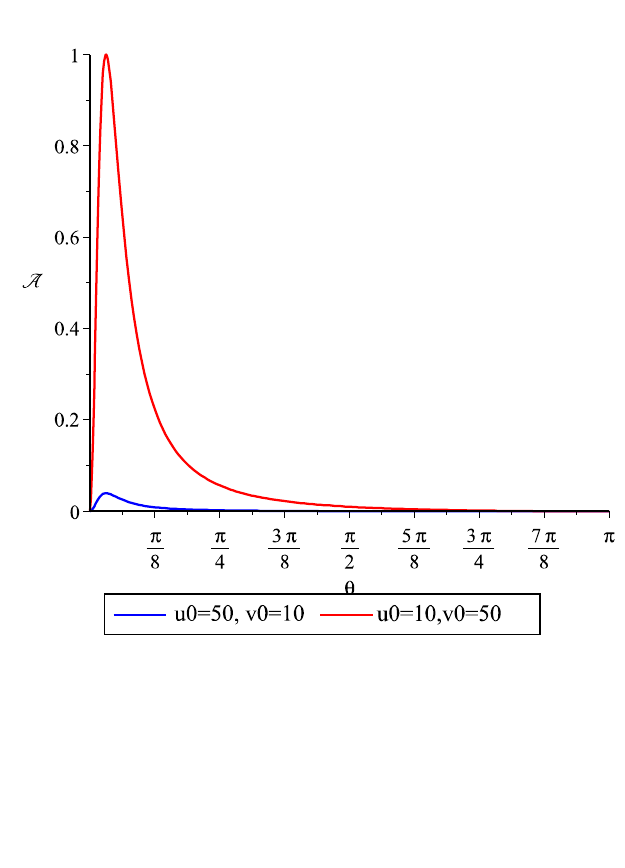} \\
\end{minipage}
\hfill
\begin{minipage}[H]{0.47\linewidth}
\centering
\includegraphics[width=1\linewidth,viewport= 60 493 240 733,clip]{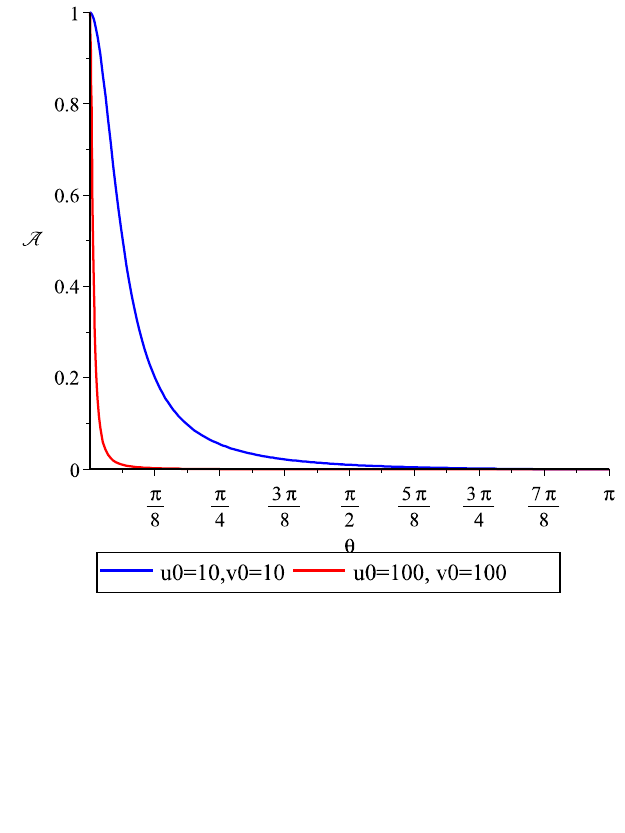}
\end{minipage} \\
\vskip-80pt
\caption{The dependence of the total amplitude of the spin-flip transition on the angle
$\vartheta$}
\label{fig1}
\end{figure}

The spin-flip probability $W=W_2+W_4$ is the probability to find the neutrino in a state with the helicity different from the initial one. The spin-flip probability depends on four frequencies, and it is restricted from above by the total amplitude $\mathpzc{A}$, which does not depend on the number density of the background fermions. It depends on the $4$-velocities of the medium $v^\mu$ and the neutrino $u^\mu$. Actually, it depends on the absolute values of the velocities and the angle $\vartheta$ between them. Fig. \ref{fig1} illustrates the typical dependence of the total amplitude on the angle $\vartheta$. If the neutrino is moving faster then the medium, then the maximum value of the total spin-flip amplitude is $ \mathpzc{A}_{max}=(v_0^2-1)/(u_0^2 -1)$. If the neutrino is moving slower then the medium, the maximum value of the total spin-flip amplitude is $ \mathpzc{A}_{max}=1$.

However, the maximum of the total amplitude does not always coincide with the maximum of the probability. For example, when the $4$-velocities of the neutrino and the medium are exactly equal ($\vartheta=0$), the amplitude is equal to unity, but as the relative velocity of the neutrino and the medium is equal to zero, there are no spin-flip transitions in this case.

\begin{figure}
\begin{minipage}[H!]{0.47\linewidth}
\centering
\includegraphics[width=\linewidth]{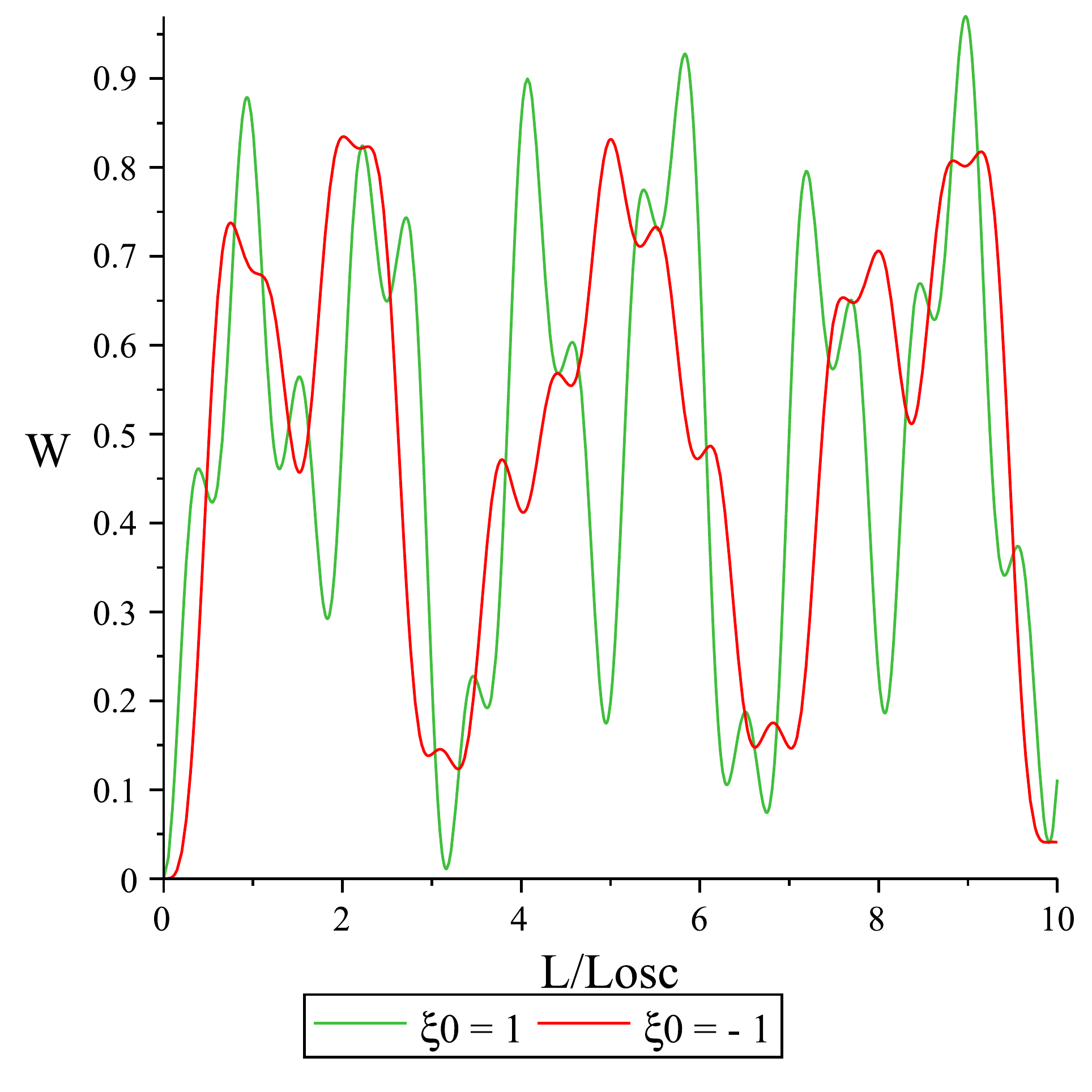}
 a) $u_0=10$, $v_0=50$,
 $k=10$
\end{minipage}
\hfill
\begin{minipage}[H!]{0.47\linewidth}
\centering
\includegraphics[width=\linewidth]{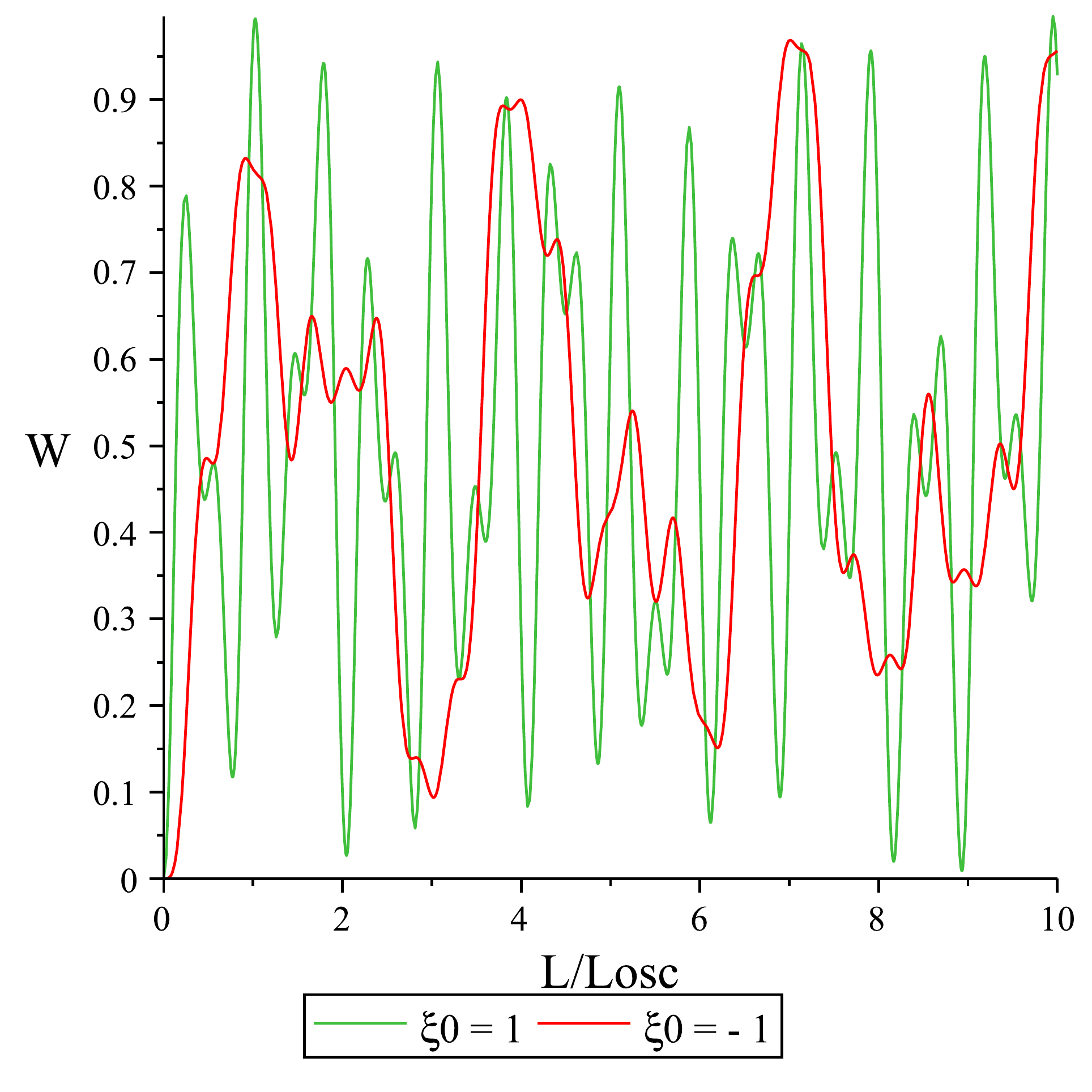}

b) $u_0=10$, $v_0=50$,
$k=20$
\end{minipage}
\caption{The dependence of the spin-flip probabilty on the distance between the source and the detector for $u_0=10$, $v_0=50$, $\cos{\vartheta}=\cos{\vartheta_{max}}$}
\label{fig_2}
\end{figure}

\begin{figure}
\begin{minipage}[H!]{0.47\linewidth}
\centering
\includegraphics[width=\linewidth]{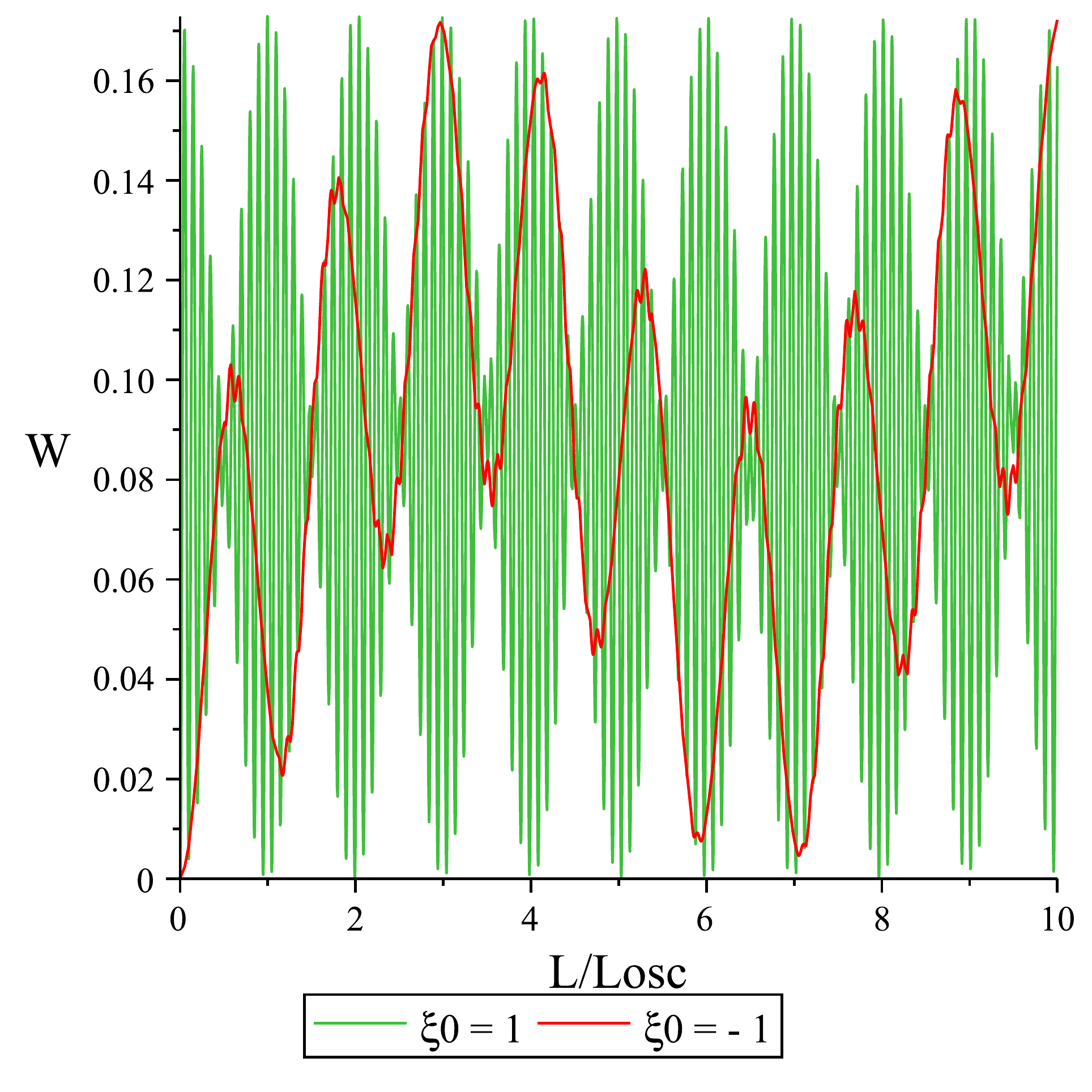}
\\ a) $u_0=10$, $v_0=50$,
 $\cos{\vartheta}=0.9$
\end{minipage}
\hfill
\begin{minipage}[H!]{0.47\linewidth}
\centering
\includegraphics[width=\linewidth]{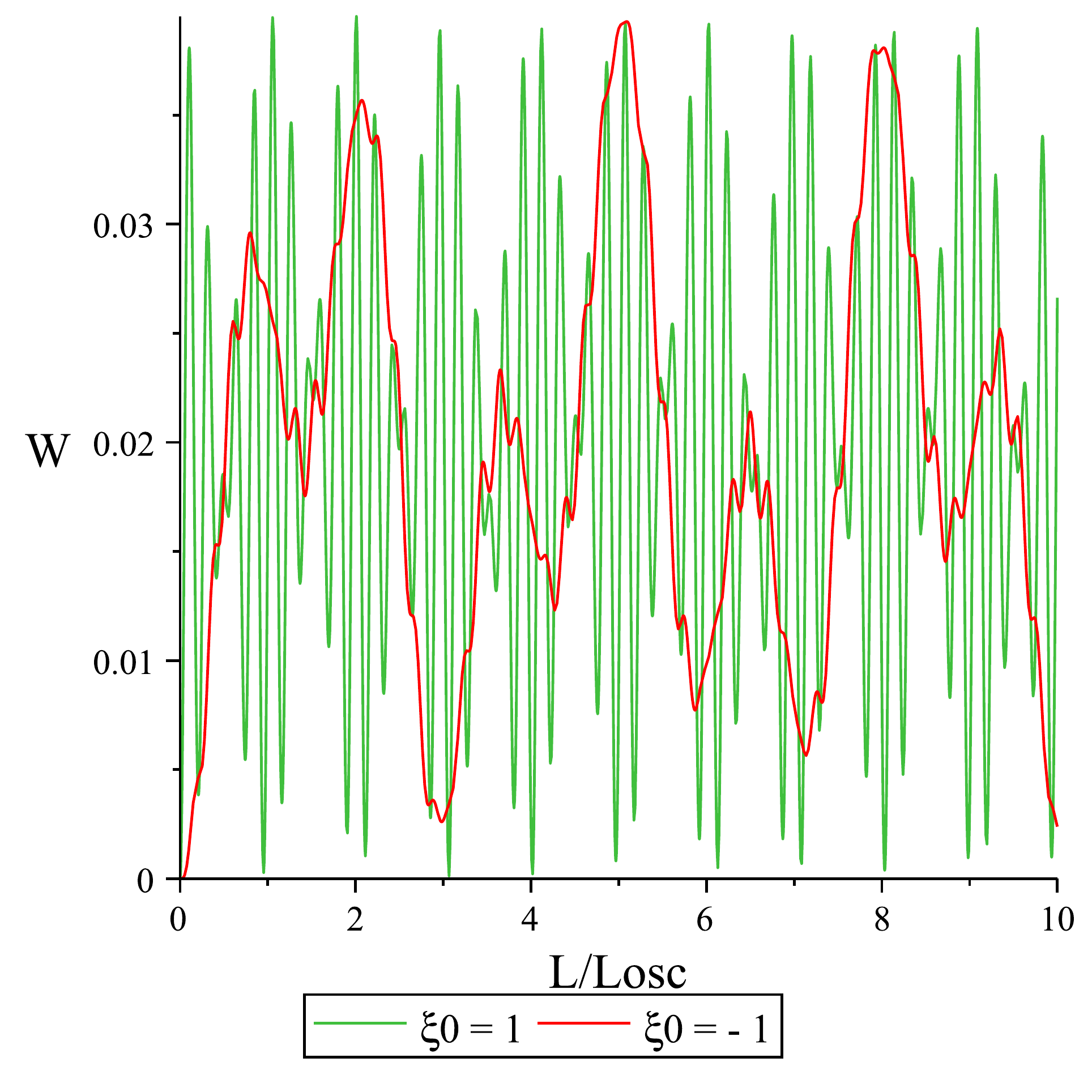}
\\ b) $u_0=50$, $v_0=10$,
 $\cos{\vartheta}=\cos{\vartheta_{max}}$
\end{minipage}
\caption{The dependence of the spin-flip probabilty on the distance between the source and the detector for $k=10$}
\label{fig_3}
\end{figure}

The influence of the medium on the neutrino propagation is characterized by the parameter $k$
\begin{equation}\label{k}
  k=\frac{\sqrt{2}G_{\mathrm F} n^{(e)}}{|m_1 - m_2|},
\end{equation}
\noindent which is proportional to the number density $n^{(e)}$ of the electrons of the medium in the laboratory frame. Fig. \ref{fig_2} and Fig. \ref{fig_3} illustrate the typical dependence of the spin-flip probability  on the distance between the source and the detector. The scale on the horizontal axis is $L/L_{osc}$, where $L_{osc}$ is the flavor oscillation length in vacuum.

Fig. \ref{fig_2} a) and Fig. \ref{fig_2} b) illustrate the dependence of the spin-flip probability on the distance in the case, when the neutrino is moving slower then the medium. The figures are given for the angle value $\vartheta_{max}$, which corresponds to the maximum of the total amplitude. The green line  corresponds to the neutrino, which  initially had the electron flavor ($\xi_0=1$) and the red line corresponds to the neutrino which initially had another flavor ($\xi_0=-1$).
It is interesting, that the difference in the spin behavior of the neutrinos, which initially had different flavors, is significant.

Fig. \ref{fig_3} a) illustrates the dependence of the spin-flip probability on the distance, when the angle $\vartheta$  is different from the maximum amplitude angle $\vartheta_{max}$. Fig. \ref{fig_3} b) illustrates the dependence of the spin-flip probability on the distance when the neutrino is moving faster then the medium, which is usually the case. In both cases the difference between the behavior of the neutrinos, which initially had different flavors, is significant.

\section{Conclusions} \label{conclusion}
The results presented in this paper may be used for a consistent quantum-field theoretical description of the neutrino evolution in matter.

The authors are grateful to A. V. Borisov, I. P. Volobuev and V. Ch. Zhukovsky for fruitful discussions. A. V. Chukhnova is also grateful to the Foundation for the advancement of theoretical physics and mathematics ``BASIS''  for  the scholarship.

\end{document}